\documentclass[12pt]{JHEP3}
\usepackage{epsfig,multicol,bbm}
\usepackage{cite}
\usepackage{amsmath,amsfonts,amssymb}
\input epsf.sty

\newcommand{\bright}{\begin{flushright}}
\newcommand{\eright}{\end{flushright}}
\newcommand{\bminip}{\begin{minipage}}
\newcommand{\eminip}{\end{minipage}}
\newcommand{\bcent}{\begin{center}}
\newcommand{\ecent}{\end{center}}

\newcommand{\WB}[1]{\overline{#1}}                                              

\def\be{\begin{equation}}
\def\ee{\end{equation}}
\def\bea{\begin{eqnarray}}
\def\eea{\end{eqnarray}}

\def\nn{\nonumber}

\title{\Huge Spinflation}

\author{Damien A.~Easson$^a$, Ruth Gregory$^{a,b}$,
David F. Mota$^c$,  Gianmassimo~Tasinato$^d$~and Ivonne Zavala$^b$ \\
${}^a$Centre for Particle Theory, Department of Mathematical Sciences,
Durham University, South Road, Durham, DH1 3LE, UK\\ 
${}^b$Institute for Particle Physics Phenomenology, Ogden Centre for
Fundamental Physics, Durham University, Durham, DH1 3LE, UK\\
${}^c$Institute for Theoretical Physics, University of Heidelberg,
69120 Heidelberg, Germany\\
${}^d$The Rudolf Peierls Centre for Theoretical Physics, Oxford University,
Oxford OX1 3NP, UK \\ \\
E-mail: \email{damien.easson@durham.ac.uk}\,, 
\email{r.a.w.gregory@durham.ac.uk}\,, 
\email{D.Mota@thphys.uni-heidelberg.de}\,,
\email{tasinato@thphys.ox.ac.uk}\,, \email{ivonne.zavala@durham.ac.uk}}

\preprint{DCPT/07/53 \\ IPPP/07/61 \\ DCPT/07/122}

\abstract{
We study the cosmological implications of including angular motion
in the DBI brane inflation scenario.
The non-canonical kinetic terms of the  Dirac-Born-Infeld action
give an interesting alternative to slow roll inflation, and 
cycling branes can drive periods of accelerated expansion in the Universe. 
We present explicit numerical solutions demonstrating brane inflation in
the Klebanov-Strassler throat. We find that demanding sufficient 
inflation takes place in the throat is in conflict 
with keeping the brane's total energy low enough so that local
gravitational backreaction on the Calabi-Yau manifold can be safely ignored. 
We deduce that \emph{spinflation} (brane inflation with angular momentum) can 
ease this tension by providing extra e-foldings at the start 
of inflation. Cosmological expansion rapidly damps the angular momentum
causing an exit to a more conventional brane inflation scenario. 
Finally, we set up a general framework for cosmological perturbation
theory in this scenario, where we have multi-field non-standard kinetic
term inflation.}

\begin{document}

\section{Introduction}

String theory, our best candidate for a theory of everything, has recently
suggested a possible fundamental origin for the inflaton field 
in the context of brane inflation~\cite{KKLMMT,brinfl, stringybr}. 
In a typical brane inflation scenario our 
three-dimensional Universe consists of
a stack of D-branes embedded in an extra dimensional 
bulk spacetime.  The identification of the
inflaton field with the position of a probe brane moving in the 
extra dimensions is a considerable step towards
addressing the problem of connecting the Inflationary Universe 
paradigm with fundamental particle physics.

The early constructions of brane inflation were somewhat {\it ad hoc},
fixing the compactification radius by hand, and using the inter-brane
Coulomb attraction to provide an inflaton potential. 
However, using the recent improved understanding of flux 
compactifications and moduli stabilization in string theory, 
a more consistent and compelling picture has emerged 
in which brane inflation is realized by a mobile D3-(or anti-D3)-brane 
in a known stabilized warped compactified background.

In a previous paper, we examined the motion of probe D3-branes 
and D3-anti-branes in
a typical warped geometry, allowing the branes to move 
freely in the angular directions~\cite{egtz}. These branes 
exhibited a rich variety of angular momentum dependent bouncing and 
cyclic trajectories which could be interpreted as cyclic cosmologies from 
the \emph{mirage} perspective. In mirage cosmology, \cite{mirage},
the cosmological expansion of the brane arises as a result of 
the brane's motion through a curved bulk spacetime.
The brane itself does not warp the background directly, but is
treated as a `probe', moving according to an effective action
appropriate to the nature of the brane and background. The mirage
approach has the distinct advantage of being exact in the 
case of codimension one cosmological branes, 
which do indeed move through warped backgrounds.  However,
the picture has shortcomings precisely because the brane
does not back-react on the geometry, nor can it in higher codimension
be associated with global effects like the domain wall. 
This lack of back-reaction makes
it rather difficult to see how to localise generic matter on the brane and
recover standard Einstein gravity at late times.

The mirage prescription has had a great impact on the
development of brane cosmology, and can be viewed as an essential first
step towards the understanding of the full brane universe.
In the case of a brane moving on a background with angular isometries, 
the addition of angular motion, with its conserved momentum 
to the probe brane has a critical effect on the 
brane cosmological models thus derived. For example, in an adS background,
a previously monotonic brane behaviour becomes dramatically altered
into a `slingshot' scenario, \cite{slingshot}, and of course as already
noted there is now the possibility of general bouncing and cyclic
cosmologies \cite{bouncing}. Since these possibilities are so radically
different to the simple radial motion brane inflation, it is vital to
understand which features are solely a consequence of the mirage 
point of view, and which are preserved once one takes the first steps 
towards a more fully self consistent gravitational picture.

Ideally, one would like to go beyond the mirage approximation
by finding a fully localised solution to the supergravity equations for the
cosmological brane moving in an (appropriately deformed) warped throat.
However, such a solution would have to reflect only the symmetries
allowed, in this case, the three-dimensional
non-compact spatial symmetry of the Universe. The internal six dimensions
strictly speaking need not have any fixed symmetries, since placing a
localised brane at an arbitrary point on the internal manifold will break
the background isometries. In the case of codimension one, this is not an issue,
as the only real degree of freedom is the scale factor of the Universe,
which (when extrapolated to the bulk) depends on only
two variables, and the system is integrable \cite{BCG}. However,
the additional codimensions, even in the absence of fluxes and 
warping, introduce new physical degrees of
freedom, which in principle could be excited in the non-supersymmetric
cosmological background. Even in the simple case of the codimension 
one Randall-Sundrum \cite{RS} scenario, cosmological solutions
have the additional degree of freedom of a bulk black hole. 

It is worth emphasizing that all known non-supersymmetric solutions
in higher codimension are obtained by imposing an ansatz for the metric
and/or bulk fields. Such ans\"atze have the effect of restricting
the degrees of freedom of the metric and bulk fields, rendering the
problem tractable. For example, an analysis of cosmological solutions
in higher codimension results in a full classification of solutions
under the assumption of separability of metric functions \cite{CG}, 
however, these most certainly are not the full set of solutions. 
In some cases, the imposition of even the mildest assumption
catastrophically restricts the solution space, as in codimension
two cosmology, where there is known to be no general FRW family \cite{Cline}
(although with Gauss-Bonnet terms the story is different \cite{BGNS}). 
On the other hand,
when dealing with an isolated brane it is not clear
one should be considering a local supergravity solution
at high curvatures.  The usual argument used in brane inflation is
that one can keep the internal metric, excise a ball around the brane and
simply replace with the metric appropriate to an isolated brane in
asymptotically flat spacetime. While we note that this could be
problematic (for example, a brane with a de Sitter worldvolume is {\it not}
asymptotically flat), we suspect that a full computation
would support the overall picture of brane inflation.

Therefore, in this paper, we adopt the usual approach in which our
Universe resides on a stack of
D3-branes embedded somewhere in the background spacetime and
the motion of an additional three-brane probe acts as the inflaton field.
We assume that the internal manifold metric is unchanged, except
possibly very close to the brane, and that the only gravitational
effect of the mobile brane is in the noncompact dimensions, where
it sources cosmological expansion. 
 We discuss the consistency of this hypothesis,
  finding for which regimes of the
  brane speed, the brane backreaction
  is reduced to a level compatible with
   our approximations. 
We allow the brane to have
spiral motion (i.e.\ movement in the internal angular
directions of the compactification \cite{shiuetal}) in the throat of
the warped space.
In this framework, we find that 
 the resulting cosmology
corresponds to an expanding universe 
 (that is, we do not find any more 
 bounces in the scale factor as in the 
  mirage approach). We  
  show that large angular momentum may
have a measurable effect on the brane inflation scenario, introducing
the notion of {\it spinflation}. In
the four-dimensional effective theory, the conserved angular momentum
may appear as a field (or several fields).
In particular, the presence of such {\it spinflatons} can
augment the inflaton field
producing a small number of additional e-foldings during
the early stages of inflation. The
angular momentum is rapidly inflated away. The presence of these fields can have
important implications for cosmological perturbations since,
non-adiabatic modes can play a non-trivial role. Precise
observable predictions for brane inflation are only possible in very
specific models~\cite{explicit}, therefore, we concentrate
on new general features induced by the addition of
angular momentum into brane inflation scenarios. Finally, we develop the
generalised theory of cosmological perturbations for scenarios involving multi-fields
with non-standard kinetic terms and apply the results to the present case.

The layout of the paper is as follows: In section \ref{sec:setup} we
introduce the set-up, establish notation and conventions,
and discuss the effect of angular momentum. In section \ref{sec:grav},
we discuss the gravitational consistency of the pseudo probe
brane approach, and present an exact solution of a brane on a
compactified spacetime at arbitrary velocity. We use this to derive
a rigorous bound on the brane speed which can be applied to
more complicated scenarios such as warped fluxed backgrounds. In 
section \ref{sec:ksthroat} we present numerical solutions for branes moving
on the Klebanov-Strassler throat (an illustrative warped background
for the deep IR), demonstrating how angular momentum or spinning can 
give extra e-foldings. In section \ref{sec:pert} we present a formal
calculation of cosmological perturbations in this DBI multi-field
inflationary scenario. Finally, 
in a complete model of inflation a mechanism to excite 
standard model degrees of freedom is required at the end
of the inflationary epoch (a so called \it reheating \rm period). 
A rigorous discussion of reheating in the context of brane 
inflation remains a challenge. Some initial progress has been 
made in~\cite{reheat,reheat1,reheat2}. Here, we provide a 
short discussion of yet another mechanism to reheat the 
Universe in the context of our model. 

\section{DBI Brane Cosmology}\label{sec:setup}

The basic set-up is as follows: we consider a 
flux compactification of type IIB 
string theory on an orientifold of a Calabi-Yau three-fold (or
an F-theory compactification of a Calabi-Yau four-fold), as 
in~\cite{gkp}. We further assume that the fluxes generate a warped throat
region in the internal space, hence we adopt a metric of the form:
\be
d s_{10}^2\,=\,h^{-1/2}(r)\,g_{\mu \nu} d x^{\mu} d x^{\nu}
\,+\, h^{1/2}(r) \, {\rm g}_{m n} d y^{m} d y^{n}\,,
\ee
where $h(r)$ is the warp factor, $g_{\mu \nu}$ is the metric in 
four dimensions, $ {\rm g}_{m n}$ is the metric of the internal 
space, and $r$ is the proper (radial) length with respect to this
internal metric (i.e.\ ${\rm g}_{rr}=1$).   We
embed a mobile D3-brane in this ten-dimensional space whose 
world-volume dynamics are described
by a Dirac-Born-Infeld (DBI) action. The D3-brane is free to move on 
the internal compact Calabi-Yau manifold, although we take the 
brane to be confined to the warped throat region. The basic setup 
is depicted in Fig.~\ref{fig:warpedCY}.
\FIGURE[ht]{\epsfig{file=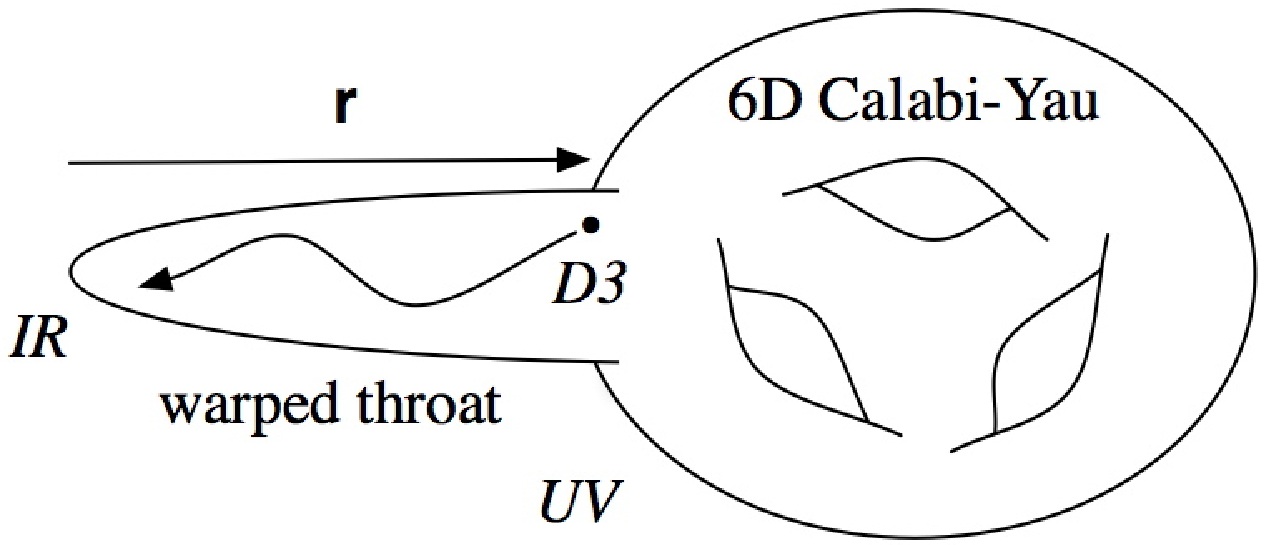,width=.8\textwidth} 
\caption{A mobile D3-brane spiraling down 
the warped throat region of a flux compactification. The
throat is smoothly glued to a Calabi-Yau threefold in the UV region. 
In principle, our Universe may exist in various parts of 
compactification, including other warped throats not shown in the
diagram. The construction involves wrapped D-branes and orientifold planes
not shown here.}
\label{fig:warpedCY}}

In our explicit computations, we take the geometry of the throat 
to be the non-singular warped deformed conifold 
(or Klebanov-Strassler) geometry~\cite{ksgeom}, the details 
of which we discuss in section~\ref{sec:ksthroat}. Mid-throat, 
the geometry is well approximated by the singular warped conifold 
(or Klebanov-Tseytlin) geometry~\cite{ktgeom}; however, we 
shall not need to make use of this simplification in our analysis.
The brane's energy-momentum produces a gravitational effect on the 
non-compact spacetime dimensions in which it is extended, as well as
a possible local distortion of spacetime. We adopt the standard 
approximation that this local distortion is negligable,
and the brane moves effectively as a test particle on the compact
manifold. It is natural to transform to field theory variables for
an effective four-dimensional description obtained by
integrating out the internal coordinates: 
\bea\label{dbigrav}
S &=& \frac{M_{Pl}^2}{2} \int{d^4x \sqrt{-g} \,R}  \nonumber \\
&& -\,g_s^{-1}\int{d^4x \sqrt{-g}  \left[   f^{-1}\sqrt{ 1
+ f\,{\rm g}_{mn}g^{\mu\nu}
\partial_\mu \phi^m\partial_\nu \phi^n} -q\,f^{-1} + V(\phi^m) \right]}\,.
\eea
The first term is the ordinary four-dimensional Einstein-Hilbert action,
which arises from dimensional reduction of the closed
string sector of the ten dimensional action,
(see discussion in section \ref{sec:ksthroat}).
The second line contains the action 
that controls the dynamics of the
fields, parameterizing the  position of the brane 
along the internal coordinates, $\phi^m = (\phi, \phi^\alpha)$ where $\phi^\alpha$ are angular
variables, and the relation 
between 
the radial geometrical and field theory variable is:
\be
\phi= \sqrt{T_3}\, r \; , \qquad 
f(\phi) = T_3^{-1} h \left (\frac{\phi}{\sqrt{T_3}} \right )
\ee 
where  $T_3= ((2\pi)^3\alpha'^2)^{-1} $ is the D-brane
tension. 
The parameter $q= 1$ ($q=- 1$) indicates that we are considering 
a brane (anti-brane). Although we include $q$ to maintain generality of
the formulae, in this paper we will only consider brane motion, i.e.\
$q=1$.   g$_{mn}$ corresponds to  
the  moduli space metric, while
$g_{\mu\nu}$ is the 4D-effective spacetime metric, that 
we take to be flat Friedmann-Robertson-Walker (FRW):
\be\label{metric}
ds^2 = -dt^2 + a^2(t) \,dx^i dx_i \,,
\ee
where $a(t)$ is the scale factor. Finally, we add a potential 
for the scalar fields $V(\phi)$ which can arise once the system is 
coupled to other sectors of the theory.  We are interested in 
the case where the potential corresponds to a mass term for the 
radially associated scalar field. In general,
the computation of the potential $V(\phi)$ is rather involved and has 
only been done explicitly in a few specific models. The potential receives
contributions from gravitational and Coulomb interactions, in addition 
to (non-negligible) correction terms from various compactification 
effects, for example, non-perturbative 
corrections to the superpotential from moduli stabilizing wrapped 
D7 (or Euclidean D3 brane instantons) \cite{KKLMMT,explicit}.
See also \cite{warpedtip} for papers in a similar framework,
\cite{wrapped} for related ideas with
wrapped branes, and \cite{Thomas:2007sj} for multiple
brane scenarios.

We assume the scalar fields $\phi^n$ are homogeneous, 
that is $\phi^n\,=\,\phi^n (t)$.
The variation of the DBI part of (\ref{dbigrav}) with 
respect to the metric gives the energy-momentum tensor.
This has the form of a perfect fluid, with energy density and pressure:
\bea\label{rhop1}
&& E=  \frac{1}{f}\left[ \gamma -q \right] +V \\
&& P=  \frac{1}{f}\left[ q- \gamma^{-1} \right] -V \label{rhop2}
\,.
\eea
Here we have defined a quantity that  plays  a key
role in the following discussion:
\be
\gamma \equiv  \frac{1}{\sqrt{1-f v^2}}
\ee
which lies in the range $\gamma\in [1,\infty)$ 
and   $v^2= {\rm g}_{mn}\dot\phi^m\dot\phi^n$.
Clearly $\gamma$ is a generalization of the usual 
relativistic Lorentz factor to the warped background.
The $(n+1)$-equations of motion for the scale
factor $a(t)$,  and the scalar fields $\phi^n$, become
\bea
&& \frac{\ddot a}{a} = -\frac{1}{6\,M_{Pl}^2g_s}\, ( E +3\,P)
\label{ray1}\\
&& \frac{1}{a^3} \frac{d}{dt}\left[ a^3 \,{\rm g}_{mn}\,\dot \phi^n 
\gamma \right]  =
\gamma\,\left(\gamma^{-1}- q \right)^2 \,\frac{\partial_m f}{2\,f^2}\, 
+\frac{\gamma}{2}
\left( \frac{\partial \,{\rm g}_{l n}}{\partial \phi^{m}}\right)  
\dot{\phi}^l \dot{\phi}^n
- \partial_m V \label{eqsfields} 
\,.
\eea
These equations are accompanied
by the Friedmann constraint
\be
H^2= \frac{1}{3\,g_s\,M_{Pl}^2} \,E = \beta E
\,,
\label{friedDBI} 
\ee
and the equation of conservation of energy:
\be
\dot E + 3H\,(E + P) =0 \label{conserva} 
\,.
\ee

We are interested in the case where
the potential is independent of the
scalar fields associated with the angular variables, so that there are conserved
angular momenta. For simplicity, we restrict to a $U(1)$
subgroup, choosing an angle which corresponds to an isometry of the metric.\footnote{Note, we will use $m,n,\dots$ 
to denote all the internal indices, and $\alpha, \beta, \dots$ to denote 
the angular internal directions.} The corresponding angular momenta, $l_\alpha$, is defined as:
%
\be \label{elem1}
\frac{1}{a^3} \frac{d}{dt} \left[ a^3\,{\rm g}_{\alpha \beta}\,\dot 
\theta^\beta \gamma\right] =0  \qquad \Rightarrow  
\qquad l_\alpha \equiv a^3\,{\rm g}_{\alpha \beta}\,\dot \theta^\beta \gamma 
\,,
\ee
The only dynamical variables that remain (to be determined
from the equations of motion) are the four-dimensional scale 
factor $a(t)$, and the position of the brane
along the radial coordinate, $\phi(t)$. In terms of the above, 
the velocity $v^2$ may be written as
\be\label{velo3}
v^2 \,=\, \frac{
\left[  \dot\phi^2 + \ell^2(\phi)/a^6
\right]}{\left[ 1+ f \,\ell^2(\phi)/a^6  \right]}
\,,
\ee
where, $\ell^2(\phi)= T_3\, {\rm g}^{\alpha \beta}\,l_\alpha l_\beta$. Therefore,
\be\label{gammal}
\gamma\,=\,\sqrt{
\frac{
1+f \,\ell^2(\phi)/a^6}{1-f \,\dot{\phi}^2}}\,.
\ee
%
%
The DBI action places a 
velocity bound on the radial field $\phi$, forcing
$1-f\,v^2 >0$. This translates  into the requirement 
$f\,\dot\phi^2 <1$, {\it independent} of
the value of  the angular momentum $\ell$ \cite{egtz}. 

An alternate way of writing equations (\ref{ray1} - \ref{friedDBI}) 
is in the first order formalism:
\begin{eqnarray}
{\dot{a}} &=&Ha\label{H1} \\
f \,\dot{\phi}^2 &=& 1-\left(1+\frac{f \ell^2(\phi)}{a^6}\right)
\cdot \left(q+f\left(\frac{ H^2}{\beta}-
V\right) \right)^{-2}\label{phidot1} \\
\dot{H}&=&-\frac{3 \beta}{2 }\left[ 2  q \left(\frac{ H^2}{\beta}-
V\right)+f \left(\frac{ H^2}{\beta} - V\right)^2
\right]\cdot  \left(q+f\left(\frac{ H^2}{\beta}
- V\right) \right)^{-1}\label{Hdot1}
\end{eqnarray}

It is often useful in 
inflationary theory to use the Hamilton-Jacobi formulation where
the dynamical variable, $H$, is written as a function of
$\phi$~\cite{HJ}. Provided the scale factor does not explicitly appear in
the main equations of motion, the Friedman and Raychaudhuri equations,
(\ref{phidot1}) and (\ref{Hdot1}), can be rewritten as a single
nonlinear first order equation for $H(\phi)$. Often, given a potential
$V(\phi)$, this leads to a direct solution for $H$. Unfortunately,
in the present case, the scale factor does appear in
(\ref{phidot1}), and therefore we cannot perform a Hamilton-Jacobi
reduction, this is because the system is now dependent not only
on the field $\phi$, but also on additional fields $\theta^\alpha$.
In addition, there is the added problem that the radial field $\phi$
is now no longer necessarily a monotonic function of time, the brane 
can, and in general does, have several oscillations up and down the
throat, thus leading to a multivalued $H(\phi)$. There is however a
way of rewriting the equations in a pseudo Hamilton-Jacobi form by 
altering our field-space coordinates
(see section \ref{sec:pert} for a full description of this).
Since the inflationary trajectory is simply a curve in the
internal Calabi-Yau manifold we can define a new coordinate, $\sigma$,
which is the proper distance along this trajectory:
\be
\sigma = \int \sqrt{{\dot\phi}^2 + T_3\, {\rm g}_{\theta\theta}\,{\dot\theta}^2}
dt \label{sigmadef}
\ee
thus $\sigma$ is a monotonic function of time, and we can now proceed with
rewriting the equations of motion along the inflationary trajectory.
In terms of this coordinate, one can readily derive $\dot\sigma$ in
terms of $H'(\sigma)$ and $f$, leading to
\be
\frac{4H^{\prime}(\sigma)^2}{9\beta^2} 
= f \left ( \frac{H^2}{\beta} - V \right )^2
+ 2 \left ( \frac{H^2}{\beta} - V \right )
\ee
Because $f$ and $V$ are both functions of $\phi$, this is not a true
differential equation for $H$, nonetheless, some useful general
qualitative information about the inflationary trajectory can be extracted. 
For example, since $|{\dot\phi}| \leq
{\dot\sigma}$, $f$ increases less rapidly, and $V$ decreases less rapidly
along an inflationary trajectory with angular momentum, which in turn
means that $H'$, and hence $\dot\sigma$ is lowered. This  feeds
into the integral for the number of e-foldings, as we show in the next 
section.

\section{Cosmology and Gravity of the brane}\label{sec:grav}

The picture outlined in the previous section is that
of a D3 brane moving on a warped background,
and driving cosmological expansion in the noncompact directions
via the energy-momentum of its motion. As yet, we have not specified
whether the brane is slowly or relativistically rolling, however,
one simple fact is clear from (\ref{friedDBI}): the 
four-dimensional scale factor is always expanding. This is because
the scale factor refers to the metric on the noncompact spacetime,
and not the induced metric on the brane. The expansion is driven by
the brane moving up {\it and} down along (damped) 
cyclic trajectories through the throat. 

In the following section, we
explore some general gravitational features of this set-up. 
Specifically, we are interested in when the brane sources accelerated 
expansion, and when the gravitational back reaction of the
brane becomes significant.

\subsection{Brane (Sp)Inflation}

The acceleration is quantified by the ``acceleration" parameter:
\be
\varepsilon\,\equiv\,-\frac{\dot{H}}{H^2} \nn
\ee
and the acceleration equation may be written as: 
\be
\frac{\ddot a}{a} = H^2 (1-\varepsilon) \,.
\ee
To achieve accelerated expansion of the scale factor $a$, $\varepsilon$ must be smaller than unity.

Using equations  (\ref{phidot1}) and (\ref{Hdot1}),
the acceleration parameter may be expressed as
\be
\varepsilon\,=\,\frac{3\beta}{2 H^2}
\left\{
\left[q+f \left( \frac{3 H^2}{\beta} -V  \right)
\right]\,\dot{\phi}^2+\frac{l^2(\phi)}{a^6}
\left[ 
q+f \left( \frac{3 H^2}{\beta} - V \right)
\right]^{-1} \right\}\label{epsilonpar}
\,.
\ee
Thus, the angular momentum plays a significant role
in determining the degree of acceleration of the cosmological expansion. 
Naively, the angular momentum appears
to hinder acceleration (because it provides a positive contribution
to $\varepsilon$~(\ref{epsilonpar})); however, the angular momentum 
{\it reduces} the brane speed $\dot\phi$ (see eq.~\ref{phidot1}) acting
to decrease the value of the first term in (\ref{epsilonpar}). 
Therefore, there is a nontrivial  interplay between the $\dot\phi$ 
and $\ell$ terms.
We shall demonstrate that the suppression of  $\dot\phi$ 
will typically win out, and the acceleration parameter $\varepsilon$ 
is effectively reduced by the angular momentum, yielding 
{\it spinflation}. 
The total number of e-foldings, $N_e$ is given by:
\be\label{nefold}
N_e(\phi) = \int H dt = \int_{\phi_i}^{\phi} 
\frac{H(\phi)}{\dot \phi} d\phi =  \int_{\sigma_i}^{\sigma}
\frac{H(\sigma)}{\dot \sigma} d\sigma 
\,,
\ee
where the intermediate $\phi$ integral is understood as a sum over
intervals where $H(\phi)$ is monotonic. Although the actual
behaviour of $H(\sigma)$ is rather involved, one can see immediately
that the overall interval for the integration in the presence of angular
momentum is increased over its absence. Since the value of $\dot\sigma$
is not increased by angular momentum without a corresponding increase in
$H$, we see that this increase in integration interval will translate 
to an increase in $N_e$.

Here we briefly summarize the different ways cosmological 
acceleration can arise, before presenting 
explicit numerical examples of these possibilities.
Note that $\varepsilon$ can be written as
\be
\varepsilon = \frac{3}{2} \frac{E+P}{E} = \frac{3}{2}
\frac{\gamma f v^2}{\gamma - 1 + fV}
\,.
\ee

In ordinary slow-roll acceleration 
$\varepsilon$ is small because $f v^2 \ll 1$, i.e.\ the field is moving
slowly. This is the most well studied model of stringy brane
inflation, and generally requires a flat, dominant, potential. 
See \cite{explicit} for interesting recent work exploring the 
validity of this set-up.

In D-acceleration  it is assumed
that the field is `rapidly' moving,  $f v^2 \simeq 1$, and
$\gamma \gg 1$, but that $fV \gg \gamma$, as in the set-up of
\cite{st}. This effect is caused by the large warp factor, rather than
a large bare velocity, and again, in past investigations
there is no angular momentum $\ell=0$. 
As we have shown in our discussion of Eq.~(\ref{epsilonpar}), 
the angular momentum plays an active role in determining the degree 
of acceleration of the system. Aside from decreasing the value 
of the slow-roll paramter $\varepsilon$, angular momentum 
contributes to the bouncing and cyclic features of  the brane 
trajectories~\cite{egtz}.  
    
In the  proximity of a turning point in the
brane trajectory,  when $\dot{\phi} \to 0$,
the size of $\varepsilon$
is reduced because  the $\dot{\phi}^2$ contribution
vanishes (\ref{epsilonpar}). Only the $\ell$ contribution
remains:
\be
\varepsilon\,=\,\frac{3\beta}{2 H^2}
\frac{\ell^2(\phi)}{a^6}\,
\left[  1+h \left( \frac{3 H^2}{\beta} - V \right)
\right]^{-1}
\label{epsilonpar2}
\,.
\ee
The obvious bouncing points occur whenever the brane 
passes through the tip of the KS geometry and
then is pulled back down by the force generated 
from the effective potential $V$.  During each pass up and down the
throat an accelerating phase is induced (whenever $\dot \phi \rightarrow 0$)
leading to a multi-inflationary type of scenario \cite{multiinfl}.
This occurs independently from
motion in the angular directions. Such oscillations, 
however, are heavily damped and it is not possible to gain
significant additional e-foldings from these short bursts of acceleration.
 
In principle, angular momentum can induce new bouncing points \emph{before} 
the brane reaches the tip of the conifold, resulting in cyclic 
trajectories~\cite{egtz}. One may  therefore hope that 
such an angular momentum induced bounce could give rise to a prolonged 
accelerating period. Unfortunately, it is not possible 
to have sufficient angular momentum to create such 
a bounce while simultaneously
generating enough e-foldings to solve the problems 
of the Standard Big-Bang model. In a successful 
inflationary model, the angular momentum is inflated away 
too quickly\footnote{See the online talk {\it 
http://online.itp.ucsb.edu/online/strings03/quevedo/} for 
a discussion of this effect on a similar configuration.} 
to induce the desired bounce before the brane 
reaches the tip of the KS geometry. Recall, from 
Eq.~(\ref{epsilonpar}) that the angular momentum contribution 
dilutes with the scale factor as $a^{-6}$. During prolonged accelerated expansion,
the angular momentum is inflated away faster than most 
other matter which dilute as
\be\label{eosl}
\rho_i \sim a^{- 3(1+ w_i)}
\,,
\ee
where the $w_i$ are the effective equation of state 
parameters for the various matter sources designated by the index $i$.
 
\subsection{Gravitational back reaction}

A concern with the pseudo probe brane approximation
for DBI inflation is that the local gravitational effect of the
brane on the throat is not generally considered. One point of view is that
this effect is extremely localized, and may be approximated by 
gluing the asymptotically flat solution in along a
tube surrounding the brane worldline. For small velocities, this
should be a good approximation, but in DBI-inflation, the brane is 
typically moving relativistically, and one must check the
corresponding increase in proper energy does not ruin this picture.
Silverstein and Tong \cite{st} gave two separate estimates 
of this effect, with two different bounds. Here we show how to
compute this bound precisely.

In order to investigate the back reaction we consider the
exact solution of the D3-brane on the compact manifold $S^1$:
\be
ds^2 = h(r,x)^{-1/2} \left [ -dt^2 + d{\bf X}^2 \right ]
+ h(r,x)^{1/2} \left [ dr^2 + r^2 d\Omega_{_{\rm IV}}^2 + dx^2 \right ]
\ee
where $x$ has periodicity $2\pi R$, and $h = 1+{\cal H}(r,x)$, with
\be
{\cal H}(r,x) = \frac{\pi g_s \alpha^{\prime 2}}{R r^3} \left \{
\frac{r}{R} \frac{\cosh(r/R) \cos(x/R) -1}{(\cosh(r/R)-\cos(x/R))^2}
+\frac{\sinh(r/R)}{\cosh(r/R)-\cos(x/R)} \right \}
\,.
\ee
This is a different physical set-up to the CY 
internal manifold, however, it is similar in the sense that the
brane is moving in a compact direction, and crucially, allows
for an exact computation of the gravitational field.

It is now simple to explore the effect of motion on the internal 
manifold: one performs a lorentz boost before making the 
identification. This has the effect of replacing the time and
internal coordinate with the new coordinates
\be
t(\tau,\xi) = \gamma (\tau+v\xi) \;\;, \qquad
x(\tau,\xi) = \gamma (\xi+v\tau)
\ee
thus rendering the metric explicitly time dependent. In the new 
coordinates:
\be
ds_{\tau,\xi} = -h^{-1/2}(1-\gamma^2 {\cal H}v^2) d\tau^2
+ 2v {\cal H} h^{-1/2} \gamma^2 d\tau d\xi + 
h^{-1/2} (1+\gamma^2 {\cal H}) d\xi^2
\ee
Demanding, that at $r=0$, the effect of the brane is negligable
at the scale of the compactification implies 
\be
\frac{4\pi g_s \alpha^{\prime 2}}{R^4} \gamma^2 \ll 1
\ee
On the other hand, asymptotically, the Newtonian potential can
be read off as 
\be
g_{\tau\tau} \sim 1 - \frac{\pi g_s \alpha^{\prime 2}\gamma^2}{R r^3} 
= 1- \frac{3\pi G_6 M_{ADM}}{2\pi^2 r^3}
\ee
where $M_{ADM}$ is the ADM mass of the brane. Now we use $G_7=2\pi R G_6$,
and $G_7 M_{ADM} = G_{10}\mu/g_s$, where $\mu$ is the proper energy per unit
worldvolume of the D3 to obtain:
\be
\frac{3\kappa_{10}^2 \mu}{16\pi^3 g_s R^4} \ll 1
\ee
This is an exact relation for the single D3 moving and localized on an $S^1$.
Now we apply this to the D3 in the warped throat, where for the proper density
of the probe we have $\mu = T_3 f E$ in field theory variables, 
and $R$ now refers to the local ambient curvature of the throat.
This gives a bound 
\be
\gamma(\phi) - 1 \ll \frac{R^4}{l_s^4 g_s} \,,
\label{brbound}
\ee
in agreement with the second of Silverstein and Tong's bounds \cite{st}.
In other words, since $g_s \ll 1$, and $R\gg1$ for the supergravity
description to be valid, the probe brane can move hyperrelativistically
before local gravitational backreaction would seem to be an issue. 
Nevertheless, we typically find that $\gamma$ grows to large values 
extremely quickly, making this bound difficult to satisfy in a
successful inflationary scenario.

\section{The Klebanov-Strassler Throat}\label{sec:ksthroat}

The Klebanov-Strassler (KS) throat is an exact nonsingular supergravity
solution with fluxes sourced by D3 and wrapped D5 branes. We use this
explicit warped geometry in order to explore the behaviour
of DBI-inflation in the deep infrared region. 
In fact, a great deal of interesting cosmological behaviour
occurs in regions where we cannot use a pure AdS or Klebanov-Tseytlin 
approximation to the geometry.  

The background geometry has the form \cite{ksgeom}: 
\be 
ds^2 =
h^{-1/2} g_{\mu\nu} \,dx^\mu dx^\nu + h^{1/2} \,ds^2_6 \ee 
where 
\bea\label{KS2} ds^2_6 &=&
\frac{\epsilon^{4/3}}{2}\,K(\eta)
\Bigg[\frac{1}{3\,K(\eta)^3}\,\left\{d\eta^2 + (g^5)^2\right\} \nn \\
&& \hskip1cm + \cosh^2{(\eta/2)} \left\{(g^3)^2 +(g^4)^2\right\}
+\sinh^2{(\eta/2)} \left\{(g^1)^2 +(g^2)^2 \right\} \Bigg]\,, 
\eea
and
\bea
K(\eta) &=& \frac{(\sinh{(2\eta)} - 2\,\eta)^{1/3}}{2^{1/3} \sinh{\eta}} \\
h(\eta) &=& 2^{2/3}(g_s M\alpha')^2 \epsilon^{-8/3} \int_\eta^\infty{dx 
\frac{x\,\coth{x} -1}{\sinh^2{x}} (\sinh{x}-2x)^{1/3}} 
\label{warpfa}
\,.
\eea
Here, $M$ quantifies the amount of three-form flux present in the 
background (for details see \cite{ksgeom}), the parameter
$\epsilon^{2/3}$ has dimensions of length and defines the UV scale
of the throat \cite{hko}, and the explicit form of the one-form 
basis $\{g^i\}$ (given in \cite{ksgeom,egtz}) is not 
important for our present investigation. The metric (\ref{KS2}) 
has angular isometries, and therefore 
conserved charges are present. For concreteness, we take the motion
to be in the angular variable which contributes identically 
to $g^2$ and $g^4$. The warp factor for KS (\ref{warpfa}) is 
shown in  Fig.~\ref{fig:hKS}.

\FIGURE[ht]{\epsfig{file=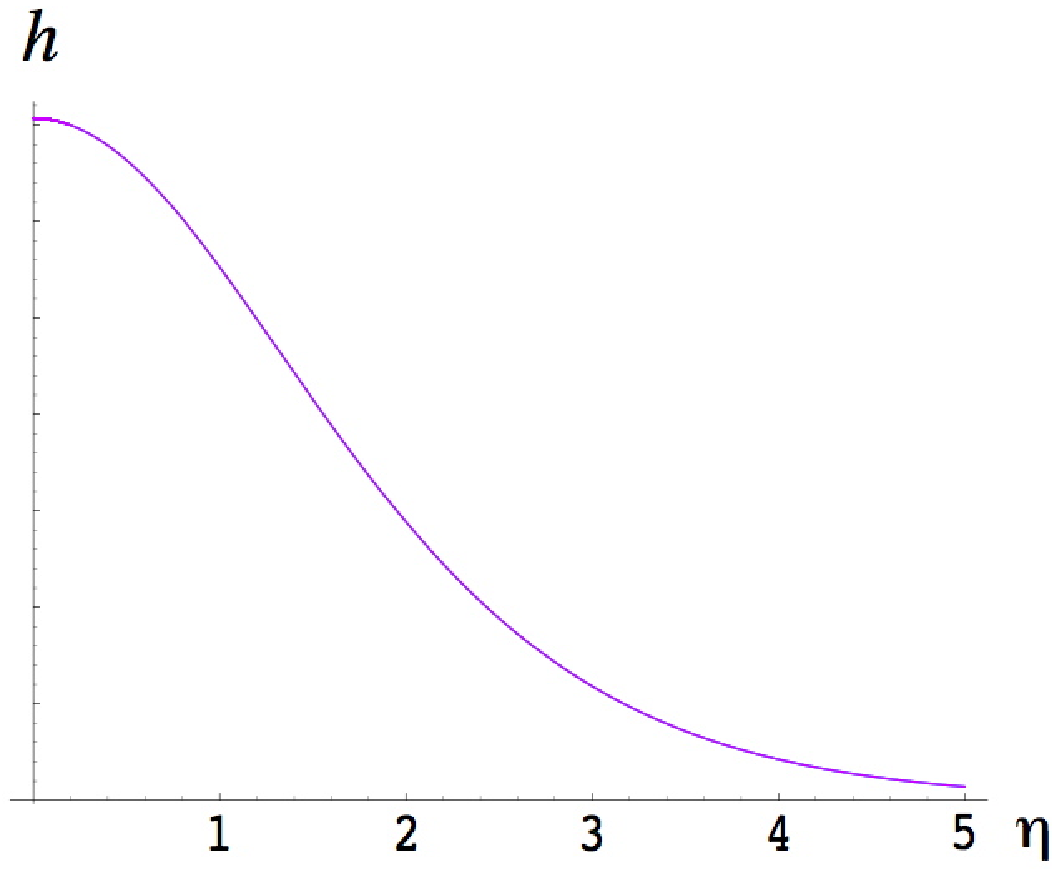,width=.7\textwidth} 
\caption{Warp factor for the KS geometry.}%
\label{fig:hKS}}

Before discussing the brane motion, we estimate the ranges
for the parameters appearing in the geometry and equations of
motion. The effective four-dimensional Planck mass is obtained
from a reduction of the 10D supergravity action:

\bea
\frac{1}{\kappa_{10}^2} \int &&\! \!\!\! d^{10} X 
\sqrt{G_{10}} R_{10} \\  \nonumber
&=&
\frac{1}{\kappa_{10}^2} \int d^4x d^6 y \sqrt{{\rm det}g_{\mu\nu}}
\sqrt{{\rm det}{\rm g}_{mn}} h^{1/2} \{ h^{1/2} R(g_{\mu\nu}) + \dots \}
\eea
Thus the Planck scale is given by
\be
M_{Pl}^2 = \frac{V_6^{(w)}}{\kappa_{10}^2} 
= \frac{{\mathcal V}_6}{(2\pi)^6\pi g_s^2} M_s^2 
\ee
where
\be
V_6^{(w)}  = \int d^6y \,\sqrt{g_6}\,h > \frac{\sqrt{2}\epsilon^4\pi^3}{3} 
\int_0^{\eta_{_{UV}}} h(\eta) \sinh^2\eta \ d\eta
\ee
is the weighted volume of the Calabi-Yau. We have also
used $\kappa^2_{10} = (2\pi)^7\,\alpha'^4\, g_s^2/2$, and
expressed the Planck mass in terms of the string scale
$M_s^{-2} = \ell_s^2 = \alpha'$, and the volume in string units
$ {\mathcal V}_6 \equiv {V_6}/\ell_s^6$. 

Performing this integral for
the KS throat shows that for $\eta_{_{UV}} \sim {\cal O} (1-10)$, the volume is
well approximated by
\be
V_6^{(w)}  \simeq \frac{\epsilon^{4/3} \pi^3}{3} (g_sM\alpha')^2
\eta_{_{UV}}^3
\,.
\ee
We can understand this intuitively since the metric near the tip of the 
throat is approximately that of a three-sphere times the neighborhood
of the origin of {\bf R}$^3$. The flux parameter, $g_s M$, is related to $M_{Pl}$, by
\be
g_s M < \left ( \frac{4}{\eta_{_{UV}}} \right )^{3/2}
\frac{\ell_s}{\epsilon^{2/3}} \sqrt{3} g_s \pi^2 \frac{M_{pl}}{M_s}
\,.
\ee
Thus in order to have large $g_sM$, which we require for
the supergravity approximation to be valid, we must take either
$M_{Pl}$ large or $\epsilon$ small.
On the other hand, 
the parameter that measures the coupling to gravity is
\be
\beta = \frac{1}{3\,g_s\,M_{Pl}^2}
\,,
\ee 
and the bigger $\beta$, the stronger the backreaction caused 
by the moving brane, and hence the greater the inflationary
effect. However, $\beta$ cannot be too large, otherwise we 
are outside the regime of perturbative string gravity.
A reasonable choice of parameters is
$g_s \sim 0.1$ and $M_{Pl} \sim 10^2$. With this choice of 
parameters and an $\eta_{UV} \sim 10$, we see that 
$g_s M \simeq 40-50 \epsilon^{-2/3}$.

\subsection{Numerical solutions}

We consider the motion of a D3-brane, allowing it to spiral along 
an angular coordinate.
Transforming to field theory variables, the radial
coordinate is defined as:
\be
r_{KS} = \frac{\epsilon^{2/3}}{\sqrt{6}}
\int_0^\eta \frac{\sinh x \, dx}{(\sinh x \cosh x - x )^{1/3}}
\label{ksrdef}
\ee
with $\phi = \sqrt{T_3}\,r$, and $f = h/T_3$ as before. Note
that we now have a rather convoluted relation between the ``natural''
geometric coordinate $\eta$, in which the metric is expressible in
terms of analytic functions (or integrals), and the field theory
coordinate $\phi$, which is the canonically normalized inflation field,
and a natural coordinate to use for cosmological inflation. Most
work on brane inflation uses an internal metric in which the natural
radial coordinate coincides with the proper distance along the throat,
however, because we are neither slow-rolling, nor restricting ourselves to a
small region of the throat, we must necessarily consider the
proper  normalization and definition of the inflaton field, which
leads to the use of (\ref{ksrdef}). For the angular field, we
use the coordinate $\theta$, which remains dimensionless.

In principle, the potential $V(\phi)$ could have contributions 
from a variety of sources, so
in order to extract generic features of introducing 
angular momentum into brane inflation scenarios we
consider the simple quadratic potential, $V=m^2\,\phi^2$.
We integrated the brane equations for a variety of parameter values
and initial conditions.  The number
of e-foldings is mainly determined by $g_sM$ and $m$, and is largely
insensitive to the value of $\epsilon$.
We find that the larger the inflaton mass,
the greater the amount of inflation, 
as in the original DBI inflation model \cite{st}, with 
(roughly) $N_e\propto m g_sM$.  
We also find that the number of 
e-foldings increases the further up the throat we start the brane rolling,
as expected.
When the angular momentum is turned on, the acceleration
is increased. 
The amount of acceleration provided by the 
angular motion (although not enough to
account for the entire number of e-folds necessary to solve the horizon
problem, etc.) can contribute with a handful of e-foldings. 
Figures \ref{fig:phi} and \ref{fig:SF} show a sample inflating 
solution with $g_sM=300, \,\epsilon=0.05, \, l_\theta \sim 5.8 \times 10^6$.
\FIGURE[ht]{\epsfig{file=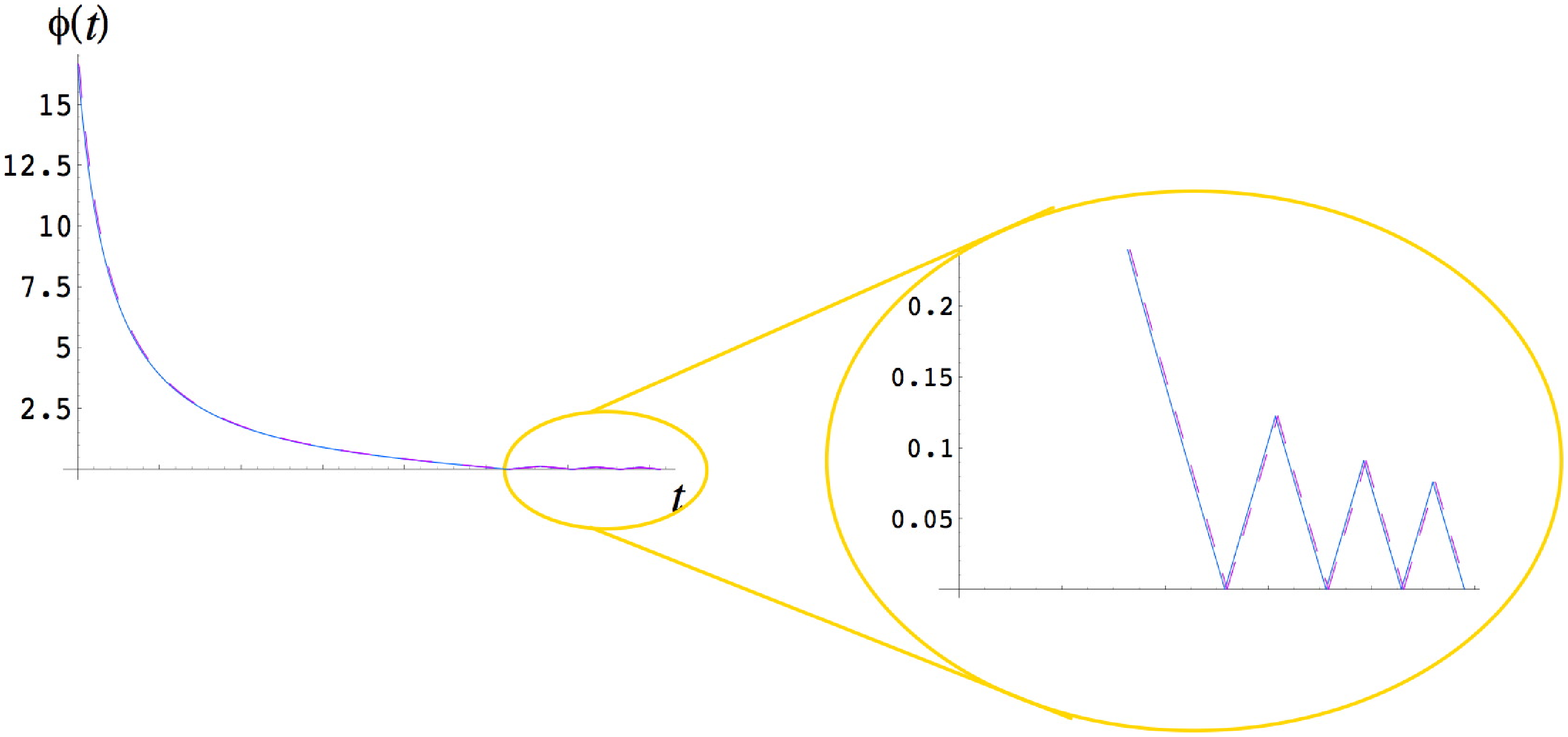,width=.9\textwidth} 
\caption{A sample plot of $\phi(t)$ for an inflating solution
with (light purple) and without (blue) angular 
momentum $l_\theta$.} 
\label{fig:phi}}
\FIGURE[ht]{\epsfig{file=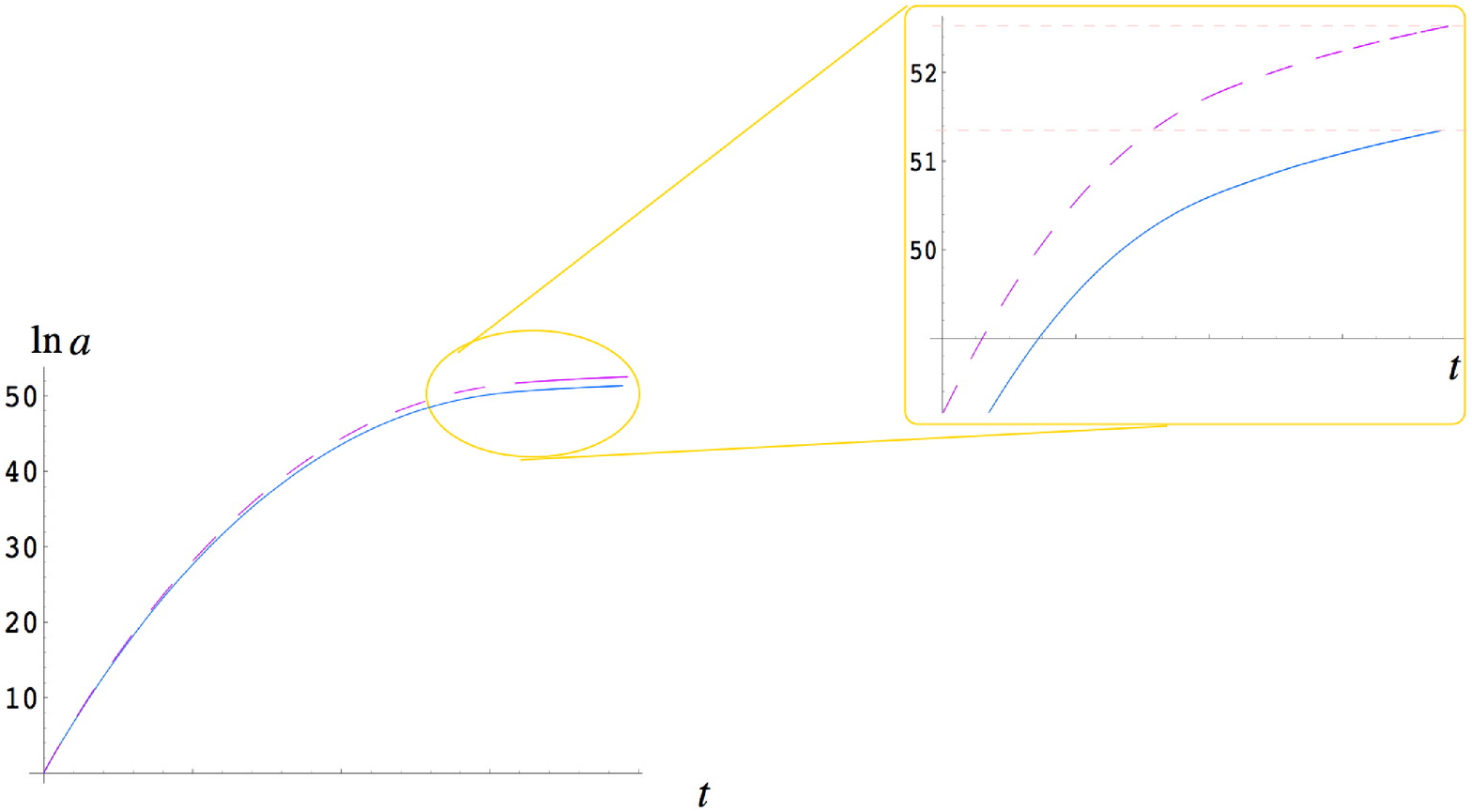,width=.9\textwidth} 
\caption{ Behaviour of the scale factor for the solution in 
figure \ref{fig:phi}.}%
\label{fig:SF}}

The solution describes a brane moving down the 
throat from the CY bulk. 
The brane accelerates as it falls, and arrives at the tip of 
the cone with non-zero velocity and continues its motion 
smoothly  back up the
throat until it reaches a turning point,  where it is 
pulled down again due to the attractive potential. 
The scale factor stops accelerating {\em before} the brane  
reaches the tip of the throat without requiring the 
presence of anti-$\WB{D3}$-branes. Therefore, inflation ends 
in a natural way, simply because of a geometric constraint. 
The KS warp factor approaches a constant value at the 
tip of the conifold. 
The brane bouncing motion continues, with the brane moving up and down 
for a few more cycles, causing the Universe to expand and accelerate briefly 
near the turning points. As the brane loses energy and reaches the minimum 
of its potential, the Universe expands as a stiff fluid-dominated 
Universe. Equation (\ref{phidot1}) implies that at the tip,
\be
H^2 \rightarrow \frac{\beta \ell_0^2}{2 a^6}
\,,
\ee
where $ \ell_0 = \ell(\phi=0)$ is the value of the angular 
momentum at the tip. The remnant energy density in the angular momentum acts as
a source that redshifts as $\rho_\ell \sim a^{-6}$, and 
therefore, has a stiff equation of state $w_{eff} = 1$ (see Eq.~(\ref{eosl})).

When the angular momentum is turned on, it provides the 
brane with an extra  kick of  acceleration ({\em spinflation}).  
This gives a couple of additional e-foldings,
although the dominant number of e-foldings 
is provided by radial motion. 

Insisting that the full e-foldings
of inflation take place in the KS throat leads to a rather artificial
set of parameter values, which induce extremely strong relativistic motion
(see figure \ref{fig:gam}). 
\FIGURE[ht]{\epsfig{file=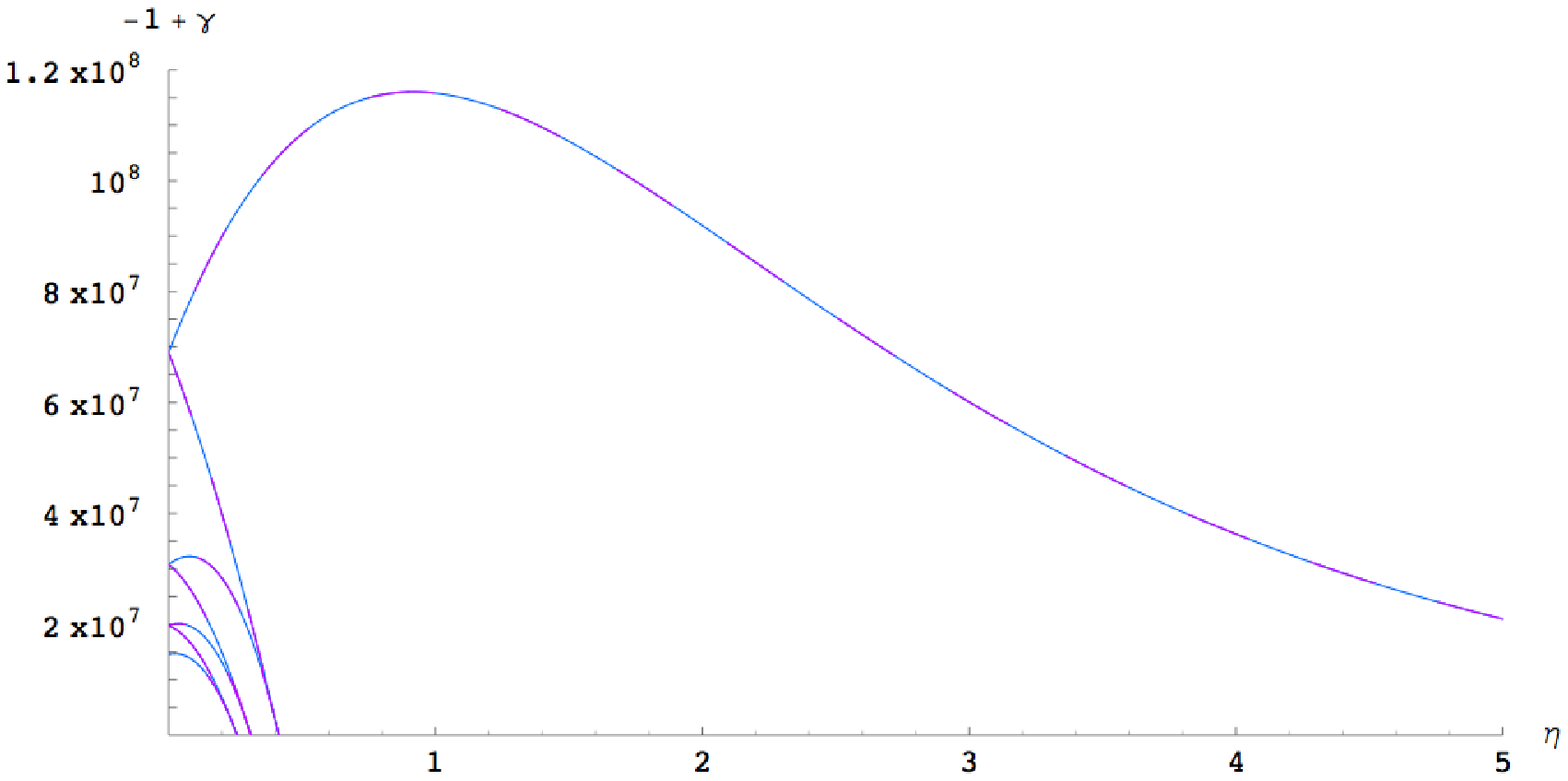,width=.9\textwidth} 
\caption{ Behaviour of the gamma factor for the solution in 
figure \ref{fig:phi}.}%
\label{fig:gam}}
To obtain sufficient e-foldings large
$g_sM$ is required, which is consistent with the supergravity approximation. On
the other hand, to keep the volume of the throat under control and
consistent with the value of $M_{Pl}$, we require that $\epsilon$ be
small, which means that the throat is very strongly warped. This
is problematic because it leads to an unacceptably large 
relativistic $\gamma$-factor, violating the back-reaction bound 
(\ref{brbound}). We were unable to find a set of parameter values 
and initial conditions which could both satisfy this bound {\it and}
supply the full e-foldings required by inflation in the KS throat. 
We suspect therefore that DBI inflation in the deep IR
is not a viable scenario,
although we have not been able to prove this result.
It is possible that a different potential or background may
evade these difficulties.

\section{Cosmological perturbations and angular motion}\label{sec:pert}

In the previous section, we found that angular momentum
acts to prolong inflation. We also noted that the main
contribution from spinning occurs at the start of inflation,
before the spinflaton is damped away. However,
even though the coherent effect of spinning may damp away,
it is possible that perturbations in the spinflaton may have
important cosmological consequences. We therefore explore 
the perturbation theory of DBI-spinflation in this section.
 
Garriga and Mukhanov, \cite{garmuk}, developed a  treatment of 
cosmological perturbations in single field inflationary models
with non-standard kinetic terms, showing that
the dynamics of linearized perturbations provides a 
spectral index for curvature perturbations very near to one.  
Building on these results, Alishahiha,
Silverstein and Tong  \cite{AST} applied their analysis to the
the D-acceleration framework, also in the single field case. 
Typically, although Gaussian perturbations
provide a spectral index compatible with observations, 
non-Gaussianities are generally non-negligible, and
provide an observational signature for these models. Interestingly, bounds
from non-Gaussianities  considerably limit the range of
the field $\phi$ \cite{AST,nongauss}.  

If angular momentum is turned on, we enter into a
multi-field inflationary scenario, and non-adiabatic
fluctuations can be generated. 
The issue of how to treat perturbations associated to non-adiabatic 
modes, in the context of DBI cosmology, has not been studied in 
full detail up to now. 
There are two key differences between this set-up and standard
slow-roll perturbation analysis. One is the non-standard DBI kinetic term,
which alters the analysis and typically gives rise to large
non-gaussianities. Another is that the geometry of the
spinflaton target space (the KS or other warped throat) is strongly curved;
even aside from the warp factor, $h(\eta)$, the intrinsic metric
${\rm g}_{mn}$ is not flat. Clearly, in slow-roll
brane inflationary models this curvature is not important, as all manifolds
are locally flat by definition. However, in
``fast-roll'' inflation (i.e.\ $f v^2 \gg 1$) the curvature of the
target space becomes important.  We emphasize that this is the
first model in which the spinflaton manifold is not flat, or conformally
flat. This complicates the perturbation analysis considerably. 
Our focus is to set up the formalism, and extract general properties 
for multi-field inflationary models with non-standard kinetic terms.  

In standard, slow-roll inflation
the presence of entropy modes normally seeds curvature perturbations,  
and, due to this fact the latter are not constant, even in the long
wavelength limit. We examine the contributions of  entropy modes to 
curvature perturbations in our multi-field DBI framework, focusing
on the long wavelength limit in
order to understand whether similar features occur here, and to explore the general characteristics 
of curvature perturbations.
We show that adiabatic
and non-adiabatic perturbations move with different
speeds, commenting on possible consequences
of this fact. We also explicitly prove that all 
perturbations remain well behaved
at the brane motion turning points, suggesting that our time dependent
trajectories
are stable under perturbations at such points. 
The resulting formalism has a natural application in cosmological 
models such as the ones we are considering here. 
The analysis is completely general and has broad applications.
Our formalism extends the seminal work of \cite{wands} for the 
canonical kinetic term case (see \cite{wandsrev} for  nice reviews,
and \cite{2fieldpert} for recent applications to string and brane cosmology).

As in standard multifield inflation, it is useful to perform a 
rotation in field space in order to rewrite the scalar equations  
in a convenient way. We re-define coordinates
based on inflationary trajectories and their normals. For simplicity,
we consider the effect of only one angular degree of freedom, although
this formalism is easily generalized. 

We define the angle $\alpha$ as:
\be
\cos{\alpha}=\frac{\dot{\phi}}{\sqrt{2 X}}\hskip 1cm 
\sin{\alpha}=\frac{\sqrt{B}\dot{\theta}}{\sqrt{2 X}}\hskip 1cm 
\ee
where 
\be
X \equiv \frac{1}{2}  \left(\dot{\phi}^2
+ T_3\,{\rm g}_{\theta \theta }\,\dot{\theta}^2
\right)\equiv \frac{1}{2} \left( \dot{\phi}^2
+B\dot{\theta}^2 \right) \,.
\ee
When $\dot{\theta}=0$, the angle $\alpha$ vanishes.
We define the averaged trajectory field $\sigma$ (see \ref{sigmadef})
and cf.\ \cite{wands}):
\be
d\, \sigma \,=\, \cos{\alpha}\, d \phi
+ \sqrt{B} \sin{\alpha}\, d \theta \,,
\label{defsigma}
\ee
which is the geodesic length introduced in (\ref{sigmadef}),
and an orthogonal, entropy field $s$:
\be
d\, s \,=\,  \sqrt{B}\cos{\alpha} \, d \theta-\sin{\alpha}\, d \phi
\,.
\label{defes}
\ee
This leads to the equalities
\begin{eqnarray}
\dot{\sigma}^2&=&2 X\nonumber\\
\dot{s}&=&0
\end{eqnarray}
in exact analogy to the flat multifield case.

\FIGURE[ht]{\epsfig{file=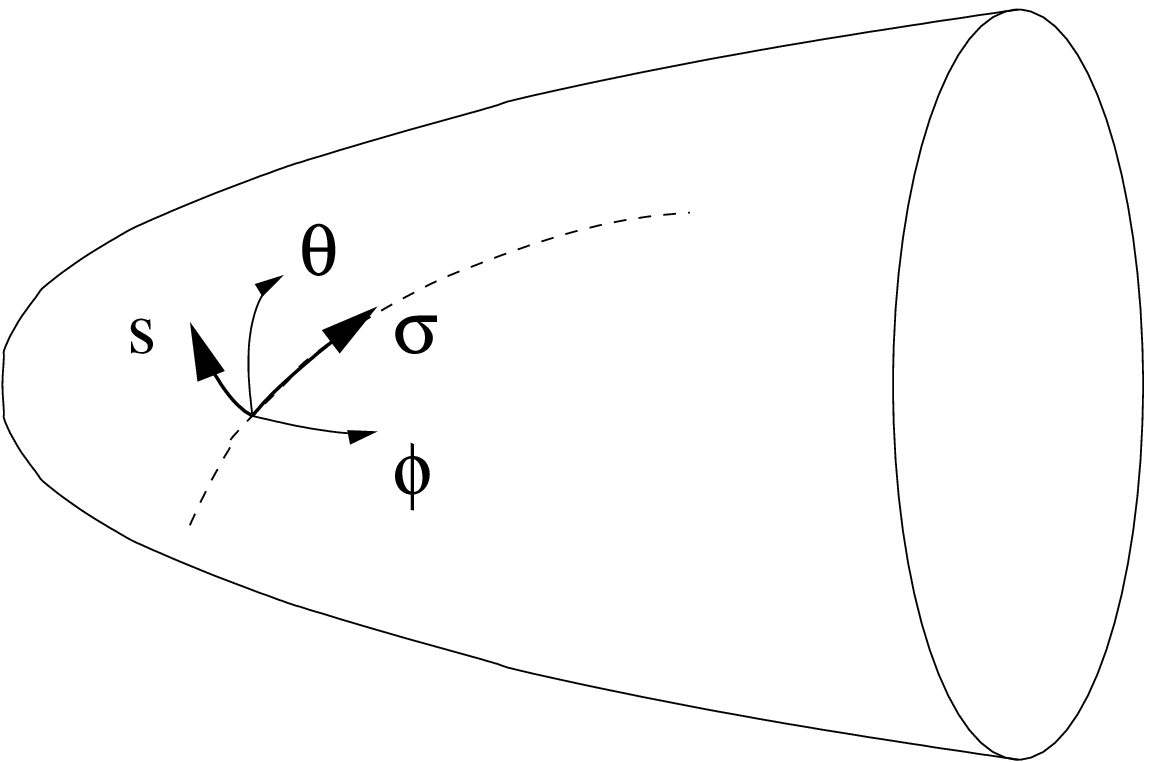,width=.7\textwidth} 
\caption{
Illustration of the multifield space.}
\label{multifield}}

The equations for $\phi$ and $\theta$ can be
rewritten in these new variables
\bea
\frac{1}{a^3} \frac{d}{d t} \left(a^3\,\gamma \,\dot{\sigma} \right)\, &=& 
- \frac{\partial V}{\partial\sigma}
+ \frac{\gamma(\gamma^{-1}-1)^2}{2f^2} \frac{\partial f}{\partial\sigma}
\label{brantra}\\
\gamma \,\dot{\sigma}\, \dot{\alpha}\,&=& - \frac{\partial V}{\partial s} +
\frac{\gamma(\gamma^{-1}-1)^2}{2f^2} \frac{\partial f}{\partial s}
+ \frac{\gamma {\dot\sigma}^2}{2B} \frac{\partial B}{\partial s}
\label{dimalfa} 
\eea
These  expressions  show that $\sigma$ can be regarded as an averaged
brane trajectory: it satisfies a dynamical equation,
(\ref{brantra}), typical of dynamical evolution under a potential,
with modifications from the warped geometry (see Figure 6).
Equation (\ref{dimalfa}) is first order, and describes 
the dynamics of the angular field $\alpha$, that is responsible for 
the departure of the brane trajectory from the radial motion.
Note that conservation of the angular momentum gives the relation
\be
\sin\alpha = \frac{l}{a^3 \gamma \dot{\sigma}\sqrt{B}}\label{solforalfa}
\ee

Now consider linear perturbation theory. We express the perturbations 
for the scalar fields in terms of perturbations in these new
variables:
\begin{eqnarray}
\frac{\delta \sigma}{\dot{\sigma}}&\equiv&
\cos^{2}{\alpha}\,\left(
\frac{\delta \phi}{\dot{\phi}}
\right)+
\sin^{2}{\alpha}\, \left(
\frac{\delta \theta}{\dot{\theta}}
\right)\,\\
\frac{\delta s}{\dot{\sigma}\,\sin{\alpha} \cos{\alpha}}&\equiv&
\frac{\delta \theta}{\dot{\theta}}-
\frac{\delta \phi}{\dot{\phi}}
\,.
\end{eqnarray}
Since the spatial part of the energy-momentum
perturbation is proportional to the metric, the scalar 
metric perturbations can be written in longitudinal 
gauge as \cite{MFB}: 
\be
d s^2\,=\,-\left(1+2\Phi\right) \,d t^2+\left(1-2\Phi\right) a^{2}(t)
\gamma_{ij}d  x^i d x^j \,.
\ee

Following \cite{garmuk}, we define gauge invariant 
combinations of the perturbations that simplify the equations. 
Let $\xi$ and $\zeta$ be:
\bea
\Phi a &\equiv & \frac{3\beta}{2}H\xi ,  \label{def1}\\
\cos^{2}{\alpha}\,\left( \frac{\delta \phi}{\dot{\phi}} \right)+
\sin^{2}{\alpha}\, \left( \frac{\delta \theta}{\dot{\theta}} \right)
=\frac{\delta \sigma}{\dot{\sigma}} &\equiv&
\frac{\zeta }{H}- \frac{3\beta\xi}{2a} .  \label{def2}
\eea
The perturbation $\zeta$ is usually called the curvature perturbation 
(in uniform-density hypersurfaces). The perturbation
$\xi$ corresponds to the Newtonian potential in longitudinal gauge,
while $\delta s$  is the entropy perturbation (notice that it is 
gauge invariant).  It is  common to work with the 
canonical Sasaki-Mukhanov variable used to quantise the field 
fluctuations coupled to first order metric perturbations during inflation  
\be\label{sasmuk}
\delta \sigma_{\Phi}\, \equiv \,
\frac{\dot{\sigma}}{H}\,\zeta
\ee
which corresponds to a gauge invariant perturbation for a spatially flat gauge. 
In the long wavelength limit,  it is equivalent to the curvature
perturbation.
 
We now study the three dynamical  perturbations $\zeta$, $\xi$,
and $\delta s$.  While the perturbations $\zeta$ and $\xi$ are
familiar from the analysis of \cite{garmuk}, here we 
must consider entropy perturbations associated
with the generation of non-adiabatic fluctuations for the curvature
perturbations. The perturbed Einstein equations for energy and momentum
readily give 
\begin{eqnarray}
\dot{\xi} &=& \frac{a\left(E +P\right) }{H^{2}}\zeta ,  \label{pereq1}\\
(E+P)\, \frac{ \dot{\zeta}}{H} &=&
\frac{H\,c_S^2}{a^3 } \left( \Delta  \xi \right)\, -\,  \tan{\alpha}
\,\left( \dot{P}-c_S^2 \dot{E} \right)\, \frac{\delta s}{\dot{\sigma}}
\nonumber\\
&=&
\frac{H\,c_S^2}{a^3 } \left( \Delta  \xi \right)\, -\, \sin\alpha
\, \left ( - V_{,\phi}(1+c_S^2) + \frac{f_{,\phi}}{f^2} (1-\gamma^{-1})^2
\right)\, \delta s
\label{pereq2}
\end{eqnarray}

Equation (\ref{pereq2}) shows how
the entropy perturbation couples to the curvature perturbation $\zeta$,
when $\alpha \neq 0$. This induces non-adiabatic entropy modes that
do not vanish even in the long wavelength limit, $k^2\ll H^2 a^2$,
or when the potential $V$ vanishes (this is
different to the case discussed in \cite{finelli}).
Clearly, (\ref{pereq1}) and (\ref{pereq2}) can be combined to give
a second order equation, however, this is more conveniently expressed
in terms of the Sasaki-Mukhanov variable, (\ref{sasmuk}), as:
\begin{eqnarray}
\ddot{\delta \sigma_{\Phi}} + \left( 3 H
+3 \frac{\dot{\gamma}}{\gamma} \right)
\dot{\delta \sigma_{\Phi}}
&+&\left(U_{\sigma_{\Phi}}+c_s^2\,\frac{k^2}{a^2}
\right) \delta \sigma_{\Phi} \nonumber \\
\,&=&\,-\left(
\frac{H \,
\dot{\sigma} c_S^2}{a^3 \left(E+P\right)}
\right)\left[\frac{a^3 \tan{\alpha}  }{
H c_S^2}  \left( 
\dot{P}-c_S^2 \dot{E}
\right)\,\frac{\delta s}{\dot{\sigma}}\right]^{\centerdot}
\label{eqforsigphi}
\end{eqnarray}
with
\begin{eqnarray}
U_{\sigma_{\Phi}}&\equiv&
\frac{\dot{\sigma}H c_S^2}{a^3 \left(E+P\right)}\,
\left[ \left( \frac{\dot{H}}{H} - \frac{\ddot{\sigma}}{\dot{\sigma}}
\right)\,\frac{a^3 (E+P)}{\dot{\sigma}H c_S^2} \right]^{\centerdot} \nonumber
\\
&=&\left(\frac{\dot{H}}{H}-\frac{\ddot{\sigma}}{\dot{\sigma}}\right)
\left(3 H +\frac{\ddot{\sigma}}{\dot{\sigma}}+3\frac{\dot{\gamma}}
{\gamma}-
\frac{\dot{H}}{H} \right) + \left(\frac{\dot{H}}{H}
-\frac{\ddot{\sigma}}{\dot{\sigma}}\right)^{\centerdot}
\,.
\end{eqnarray}

A third, independent equation of motion that controls the evolution of 
$\delta s$ is found from the perturbed $\theta-$equation 
\be
\delta \ddot{s} \,+\, \left( 3 H +\frac{\dot{\gamma}}{\gamma} \right)
\delta \dot{s} \,+\,\left(U_{s}+\frac{k^2}{a^2} \right) \delta s 
\,=\,- \, \frac{k^2}{a^2}\,\frac{ \dot{\sigma}\,\tan{\alpha}\,
H}{a \left(E+P\right)^2}\,\left(\dot{P}- c_S^2 \dot{E}\right) \,\xi
\label{eqfors}
\ee
where 
\begin{eqnarray}
U_s\,&=&\,\tan{\alpha}\Bigg[ 3H \tan{\alpha} \left(
\frac{\ddot{\sigma}}{\dot{\sigma}}-\frac{\dot{\alpha}}{\tan{\alpha}}
+3 H c_S^2-\cos{\alpha} \frac{B'\,\dot{\sigma}}{2B} \right) \nonumber \\
&+&\dot{\sigma}\left(\frac{\dot{\alpha}}{\dot{\sigma}}
-\frac{B'\,\cos{\alpha}}{2B\,\tan{\alpha}}
\right)^{\centerdot} \Bigg] \,.
\label{Usdef}
\end{eqnarray}

Even without solving these equations, we can extract a great 
deal of useful qualitative information about the behaviour of
perturbations in multifield DBI inflation.
First, the entropy perturbation evolves
independently of the curvature perturbation at large scales 
($k^2/a^2\ll1$): it couples to the curvature perturbation
only through the  Newtonian potential $\xi$. 
(This remains true even if the potential is not angularly
invariant, see (\ref{eqforsl}).) 
Interestingly, the converse is not true: at large
scales the entropy perturbation actually
seeds curvature perturbations. Equation (\ref{eqforsigphi}) 
(or (\ref{pereq2l}) for the case of angular $V$ dependence)
contains contributions from $\delta s$ that do not 
vanish even in the long wave-length limit.

Another important observation is that curvature 
and entropy perturbations move 
with {\it different speeds}. While curvature
perturbations move with a speed of sound much smaller than
the speed of light (indeed $c_S^2 \ll 1$), the entropy 
perturbations move at light speed. 
An interpretation of this fact relies
on the observation that the speed of the perturbations
is related with the speed of the background quantity they
are associated with.  Curvature perturbations
are associated with the speed of the 
average trajectory $\dot{\sigma}$,
that enters the speed of sound as $c_S^{-2}\,=\,1-h\dot{\sigma}^2$.
Entropy perturbations refer to $\dot{s}=0$,
and the corresponding sound speed is $c_S^{-2}\,=\,1-h\dot{s}^2=1$.~\footnote{We thank Kazuya Koyama for discussions on this point.}
A consequence of this fact is  that
entropy  perturbations can re-enter the horizon much earlier than
the curvature perturbations in DBI models.  It would
be interesting to study consequences and applications
of this fact.

Finally, coupling between adiabatic and isocurvature
modes generally becomes stronger at the turning points,
if the angular momentum does not vanish. Since 
$\sin{\alpha}=1$ at these points, the coefficient of 
the term proportional to $\delta s$ in (\ref{eqforsigphi})
acquires its maximal size.
This raises an important physical and technical issue: are the
perturbations finite at turning points? We have learned that
the brane trajectory will encounter
bounces during its evolution, and this can  have interesting applications for cosmology.
In the absence of angular momentum, the entropy perturbations
decouple from curvature perturbations, however, in the presence
of angular momentum this is no longer the case, and at the bounce
when $\cos{\alpha}=0$, there is a potential divergence in this 
coupling from the $\tan\alpha$ term on the RHS of (\ref{eqfors}).
In addition, the expression for $U_s$, (\ref{Usdef}), is potentially
singular for the same reason. Clearly, such a singularity would be
catastrophic for cosmological perturbation theory, and of course
does not arise in conventional inflationary models. We must therefore
check that this is not an issue for spinflation.

Physically, we do not expect the turning point to exhibit any particular
pathology. Inflation is produced by the motion of a D3-brane on an
internal manifold, and cosmological perturbation theory is a combination
of gravitational perturbations in the noncompact manifold and
internal fluctuations in the trajectory of the brane, in other words
the brane ``flutters'' on the internal manifold as a function of the
spacetime coordinates. Picturing the moving D3-brane as a fluttering
sheet shows that although we might not expect every point on
the brane to turn at the same time, overall the brane will turn and there
should be no associated singularity. Similarly, from the 
field theory point of view, a field oscillating in a potential well experiences
several turning points, and in no way are these associated with 
singularities. Therefore, although the formulae appear ill-behaved at
$\cos{\alpha}=0$, we do not expect that there is a technical problem.

The equation  for curvature perturbations,
(\ref{pereq2}), is regular at any bounce, however, the entropy
perturbation equation is somewhat more subtle. The RHS of (\ref{eqfors})
is regular, for the same reason as (\ref{pereq2}). At first
sight (\ref{Usdef}) seems problematic, containing terms proportional
to $\tan^2\alpha$. Using (\ref{dimalfa}), the derivative
in the second term of (\ref{Usdef}) is:
\be
\dot{\sigma}\frac{d\ }{dt} \left(\frac{\dot{\alpha}}{\dot{\sigma}}
-\frac{B'\,\cos{\alpha}}{2B\,\tan{\alpha}}
\right) = - {\ddot\sigma} \sin\alpha \left (
\gamma V_{,\phi} \frac{(1+ c_S^2)}{{\dot\sigma}^2} + 
\frac{f_{,\phi}}{f} \frac{(\gamma-1)}{(\gamma+1)} \right )
\,.
\ee
Using conservation of angular momentum:
\be
\frac{\dot{\alpha}}{\tan{\alpha}} + \frac{\ddot{\sigma}}{\dot{\sigma}}
+3H + \frac{B'\, {\dot\phi}}{2B} + \frac{\dot{\gamma}}{\gamma} = 0
\ee
to substitute for $H$ in the first term of (\ref{Usdef}), and
tracking the $V$ and $f$ derivatives carefully, we find
\be
U_s = - \tan\alpha \sin\alpha {\dot\sigma} \left (
\gamma V_{,\phi} \frac{(1+ c_S^2)}{{\dot\sigma}^2} +         
\frac{f_{,\phi}}{f} \frac{(\gamma-1)}{(\gamma+1)} \right )
\left (\frac{\ddot\sigma}{\dot\sigma} + 3H c_S^2 \right )
= {\cal O}(1) \,.
\ee
Thus the perturbations are manifestly well-behaved at any bounces,
in agreement with our expectations.

Finally, in the case of conserved angular momentum,
during  inflationary expansion entropy perturbations become rapidly
decoupled from the adiabatic ones.  Consequently, they are not likely to
drastically modify the evolution of curvature perturbations. This is
particularly evident from equation (\ref{pereq2}): indeed, the angle
$\alpha$ decreases as the cube of the scale factor (see eq. (\ref{solforalfa})). During
inflation the scale factor increases rapidly and the angle
$\alpha$ shrinks to negligible size within a handful of
e-folds.  A different situation may occur when the angular momentum is
{\it not } conserved, and the potential $V$ explicitly depends on the
angular coordinates.  In Appendix A we derive the general
form of the perturbation equations.  We leave the analysis
of the imprints of entropy perturbations to inflationary parameters in
this general case for future work. 

\section{Conclusions and Outlook}

In this paper we studied various aspects of angular motion 
of branes travelling in a warped throat. Our main focus was on
the impact of angular momentum in a DBI-inflationary scenario, where
the brane is typically moving at relativistic proper velocities. We
used the KS throat as a stereotypical example to study the
cosmologies of such trajectories numerically, and
developed an analytic perturbation formalism for general multi-field 
DBI-inflation.  We find that the angular momentum enriches
the qualitative features of the brane motion.

The four-dimensional low energy cosmology is controlled by a standard
Einstein-Hilbert term for gravity, coupled to a DBI-like action for the
scalar fields that parameterise the brane position in the higher
dimensional background. The cosmological evolution is characterised by an
always increasing scale factor. Periods of accelerated expansion can
occur in various ways. In addition to standard slow-roll inflation, the
non-canonical form of the kinetic terms impose a bound on the brane
velocity. This leads to the D-acceleration scenario of
\cite{st}: in the regions where the brane speed is reduced, by means of
this bound, the potential energy dominates and drives inflation.
Short periods of acceleration can occur when the brane
trajectories have turning points: since the brane speed in the radial
direction vanishes, the total velocity is reduced and potential terms are
more likely to dominate, providing acceleration.

We took the specific example of a warped KS throat for a
numerical study of the cosmologies arising from a spinning brane.
Our investigations indicated that building a successful inflationary
model based on the DBI-action is in tension with the
constraints coming from the validity of the supergravity
approximation and fine tuning of parameters is required. 
We found that in order to have sufficient inflation, the relativistic
$\gamma$-factor was driven to extremely large values, well in excess of 
any non-gaussianity or other bounds \cite{AST,Panda:2007ie,constraints}, 
and typically at least an order
of magnitude in excess of the back-reaction bound. This would appear to
be a major challenge for the DBI-inflation scenario within string theory.
It would be interesting to investigate how generic this conclusion is. For
example, making the effective potential $V(\phi)$ steeper by using
a $\phi^4$ term seems to ameliorate the problem. We could also include
a conformal coupling term (see \cite{Lev}) or alter the background
throat geometry.

Non-standard kinetic terms reveal novel features when 
analysing Gaussian cosmological perturbations. 
Brane motion along angular trajectories, in addition to the
radial one, implies that cosmological expansion is driven by more than one
field. Moreover, the strong warping of the throat means that the 
curvature of the target space can also contribute to the perturbation
theory. Non-adiabatic modes may play an important role on the
evolution of curvature perturbations. We derive the equations for
cosmological perturbations in this set-up, 
finding that non-adiabatic entropy modes move
with different speeds with respect to adiabatic, curvature perturbations.
This may have important consequences when studying inflationary models,
for example since different perturbations enter the horizon at different
times. Also, the features of the curved target space appear in the
effective potentials that drive the perturbations, that are
sensitive to the underlining geometry.

We focused mainly on potentials with no angular dependence, since that
was the example used in the explicit numerical cosmologies. However, for 
completeness, we include in an appendix the calculations in the
presence of angular dependence.  In this situation, the
coupling between entropy and curvature modes is likely to be suppressed
during inflationary expansion. In general, one expects that wrapped branes 
used to stabilise the throat moduli induce angular dependence, and indeed
this has been demonstrated explicitly for a particular embedding of
D7-branes in the KT background \cite{explicit}, thus
any concrete model of DBI-inflationary perturbations must include such
an analysis. A complete analysis of angular dependence is clearly 
the next step in exploring general inflationary trajectories, however,
given that the number of e-foldings seems to be related to the length
of the inflationary trajectory it is quite likely that 
even in the presence of angular dependence, the number of e-foldings
will be increased.

Finally, it is possible that angular momentum may have additional
interesting consequences for reheating the Universe at the end of
inflation. In conventional brane inflationary models an anti-$\WB D$-brane is
placed at the tip of the throat. The reheating mechanism
involves tachyon condensation, when the inflaton brane reaches the
$\WB D$-brane causing an explosive reheating process.
In our case, a $\WB D$-brane is not necessary for successful reheating.

We can envision a natural mechanism for an inflationary exit, and an
alternative scenario for reheating. It is possible that damped 
brane oscillations through the tip of the
KS throat cause the three non-compact
dimensions to heat up (similar scenarios have been proposed in
\cite{reheat,reheat2}). The brane inflaton fields $\phi$ 
may have couplings to a scalar $\xi$ or
spinor $\psi$ through an interaction Lagrangian of the form
\be
{\cal L}\,=\,-\frac12 g^2\, \phi^2 \xi^2 - h\, \phi \bar{\psi} \psi\,.
\ee
Each time the brane bounces in the throat, the inflaton can decay {\it
via} the above interactions into $\xi$ and $\psi$ particles.  The total
decay rate is consequently
\be
\Gamma \,=\,\Gamma(\phi\to \xi \xi)+ \Gamma(\phi\to \psi \psi)\,.
\ee
Details of this process depend on the model and on specific features of
the compactification. 

\section*{Acknowledgements}
We would like to thank Marta G\'omez-Reino, Lef Kofman, Kazuya Koyama, 
Liam McAllister, Fernando Quevedo, Martin Schvellinger, 
Kerim Suruliz and David Wands
for useful discussions and comments.
This research was supported in part by PPARC. DE, RG and GT 
are partially supported by the EU $6^{th}$ Framework Marie Curie 
Research and Training network ``UniverseNet" (MRTN-CT-2006-035863).
DFM is supported by the Alexandre von Humboldt foundation,
GT is also supported by the EC $6^{th}$ Framework Programme Research 
and Training Network ``UniverseNet'' (MRTN-CT-2004-503369), and
IZ is supported by a Postdoctoral STFC Fellowship. 

\appendix

\section{Cosmological perturbations for an angularly
dependent potential }

We now generalize the equations that govern the cosmological perturbations
to the case where the potential depends explicitly on the angular variable $\theta$. We start by defining a combination
$Q$ that will frequently appear:
\be
Q\,\equiv\, a^3 \gamma \sqrt{B} \, \dot{\sigma}
\sin{\alpha} 
\,.
\ee
This quantity, at the background level, satisfies the relation
$$ \partial_t Q=-a^3\,\partial_{\theta} V\,$$
so that if $\partial_{\theta} V=0$, $Q$ is constant and 
defines the conserved angular momentum. In the more general case in
which $\partial_{\theta} V\neq0$, it is straightforward to obtain
the equations that determine the curvature perturbations. We find
\begin{equation}
\dot{\xi}=\frac{a\left(E +P\right) }{H^{2}}\zeta ,  \label{pereq1l}
\end{equation}
\begin{eqnarray}
(E & + & P)\frac{ \dot{\zeta}}{H} = \\ \nonumber 
&& \frac{Hc_S^2}{a^3 }
\left( \Delta  \xi \right) -\frac{ \tan{\alpha}}{\dot{\sigma}}\left[ 
\left( \dot{P}-c_S^2 \dot{E}
\right)-\left(1+c_S^2\right)\left(E+P\right)
\frac{\partial_t Q}{Q}
\right]\,\delta s
\label{pereq2l},
\end{eqnarray}
 from which a few additional steps lead to a second order equation in terms of 
the Sasaki-Mukhanov variable. 

The equation governing entropy perturbations is:
\bea
\delta \ddot{s} &+& \left( 3 H
+\frac{\dot{\gamma}}{\gamma} \right) \delta \dot{s}
\,+\,\left(U_{s}+\frac{k^2}{a^2} \right) \delta s
= \nonumber  \\ &-&  \frac{k^2}{a^2}\frac{ \dot{\sigma}\tan{\alpha}
H}{a \left(E+P\right)^2}\left[\dot{P}-
c_S^2 \dot{E}-\left(1+c_S^2\right) \left(E+P\right)
\frac{\partial_t Q}{Q} \right] \xi
\label{eqforsl}
\eea
where $U_s$ is
\begin{eqnarray}
U_s&=&\tan{\alpha}\left[ 3H \tan{\alpha} \left(
\frac{\ddot{\sigma}}{\dot{\sigma}}-\frac{\dot{\alpha}}{\tan{\alpha}}
+3 H c_S^2-\cos{\alpha} \frac{B'\,\dot{\sigma}}{2 B}
\right)+\dot{\sigma}\left(\frac{\dot{\alpha}}{\dot{\sigma}}
-\frac{B'\,\cos{\alpha}}{2 B\,\tan{\alpha}}
\right)^{\centerdot}
\right]+ \nonumber\\
&+&\tan{\alpha} \frac{\partial_t Q}{Q}\left\{
\dot{\alpha}-\frac{B'\,\cos{\alpha} \,\dot{\sigma}}{2 B\,\tan{\alpha} }
-\tan{\alpha}\left[\dot{P}-
c_S^2 \dot{E}-\left(1+c_S^2\right) \,\left(E+P\right)\,
\frac{\partial_t Q}{Q} \right]
\right\}+ \nonumber\\
&+& \frac{a^3 \dot{\sigma} \tan{\alpha}}{Q}\,\left[
\frac{\partial^2 V}{\partial \theta^2} \,\frac{B'\,\cos{\alpha}}{2 B}
\,-\,\frac{\partial^2 V}{\partial \theta  \partial \phi} \,\sin{\alpha}
\right]
\,.
\end{eqnarray}
With the aid of these equations, it is possible to study the evolution
of cosmological perturbations for a system of
two fields with DBI action in full generality.


\begin{thebibliography}{99} \itemsep -1pt
%

\bibitem{KKLMMT}
S.~Kachru, R.~Kallosh, A.~Linde, J.~M.~Maldacena,
L.~McAllister and S.~P.~Trivedi,
JCAP {\bf 0310}, 013 (2003)
[arXiv:hep-th/0308055].

\bibitem{brinfl}
G.~R.~Dvali and S.~H.~H.~Tye,
Phys.\ Lett.\ B {\bf 450}, 72 (1999)
[arXiv:hep-ph/9812483].\\
S.~H.~S.~Alexander,
Phys.\ Rev.\  D {\bf 65}, 023507 (2002)
[arXiv:hep-th/0105032].\\
C.~P.~Burgess, M.~Majumdar, D.~Nolte, F.~Quevedo, G.~Rajesh and R.~J.~Zhang,
JHEP {\bf 0107}, 047 (2001)
[arXiv:hep-th/0105204].\\
G.~Shiu and S.~H.~H.~Tye,
Phys.\ Lett.\  B {\bf 516}, 421 (2001)
[arXiv:hep-th/0106274];\\

\bibitem{stringybr}
J.~Garc\'{\i}a-Bellido, R.~Rabad\'an and F.~Zamora,
JHEP {\bf 0201}, 036 (2002)
[arXiv:hep-th/0112147];\\
K.~Dasgupta, C.~Herdeiro, S.~Hirano and R.~Kallosh,
Phys.\ Rev.\  D {\bf 65}, 126002 (2002)
[arXiv:hep-th/0203019];\\
N.~T.~Jones, H.~Stoica and S.~H.~H.~Tye,
JHEP {\bf 0207}, 051 (2002)
[arXiv:hep-th/0203163]; \\
M.~G\'omez-Reino and I.~Zavala,
JHEP {\bf 0209}, 020 (2002)
[arXiv:hep-th/0207278],

\bibitem{egtz}
D.~Easson, R.~Gregory, G.~Tasinato and I.~Zavala,
JHEP {\bf 0704}, 026 (2007)
[arXiv:hep-th/0701252].
  
\bibitem{slingshot}
C.~Germani, N.~E.~Grandi and A.~Kehagias,
``A stringy alternative to inflation: The cosmological slingshot scenario,''
arXiv:hep-th/0611246.\\
R.~Brandenberger, H.~Firouzjahi and O.~Saremi,
``Cosmological Perturbations on a Bouncing Brane,''
arXiv:0707.4181 [hep-th].

\bibitem{bouncing}
S.~Kachru and L.~McAllister,
JHEP {\bf 0303}, 018 (2003)
[arXiv:hep-th/0205209]; \\
  C.~P.~Burgess, P.~Martineau, F.~Quevedo and R.~Rabad\'an,
  JHEP {\bf 0306}, 037 (2003)
  [arXiv:hep-th/0303170]; \\
  C.~P.~Burgess, F.~Quevedo, R.~Rabad\'an, G.~Tasinato and I.~Zavala,
  JCAP {\bf 0402} (2004) 008
  [arXiv:hep-th/0310122].

\bibitem{mirage}
A.~Kehagias and E.~Kiritsis,
JHEP {\bf 9911}, 022 (1999)
[arXiv:hep-th/9910174].

\bibitem{BCG}
P.~Bowcock, C.~Charmousis and R.~Gregory,
Class.\ Quant.\ Grav.\  {\bf 17}, 4745 (2000)
[arXiv:hep-th/0007177].

\bibitem{RS}
L.~Randall and R.~Sundrum,
Phys.\ Rev.\ Lett.\  {\bf 83}, 4690 (1999)
[arXiv:hep-th/9906064].

\bibitem{CG}
C.~Charmousis and R.~Gregory,
Class.\ Quant.\ Grav.\  {\bf 21}, 527 (2004)
[arXiv:gr-qc/0306069].

\bibitem{Cline}
J.~M.~Cline, J.~Descheneau, M.~Giovannini and J.~Vinet,
JHEP {\bf 0306}, 048 (2003)
[arXiv:hep-th/0304147].

\bibitem{BGNS}
O.~Corradini, A.~Iglesias, Z.~Kakushadze and P.~Langfelder,
Phys.\ Lett.\  B {\bf 521}, 96 (2001)
[arXiv:hep-th/0108055];\\
P.~Bostock, R.~Gregory, I.~Navarro and J.~Santiago,
Phys.\ Rev.\ Lett.\  {\bf 92}, 221601 (2004)
[arXiv:hep-th/0311074];\\
H.~M.~Lee and G.~Tasinato,
JCAP {\bf 0404} (2004) 009
[arXiv:hep-th/0401221].

\bibitem{shiuetal}
O.~DeWolfe, L.~McAllister, G.~Shiu and B.~Underwood,
``D3-brane Vacua in Stabilized Compactifications,''
arXiv:hep-th/0703088.

\bibitem{explicit}
D.~Baumann, A.~Dymarsky, I.~R.~Klebanov, L.~McAllister and P.~J.~Steinhardt,
``A Delicate Universe,''
arXiv:0705.3837 [hep-th];\\
D.~Baumann, A.~Dymarsky, I.~R.~Klebanov and L.~McAllister,
``Towards an Explicit Model of D-brane Inflation,''
arXiv:0706.0360 [hep-th].

\bibitem{reheat}
J.~H.~Brodie and D.~A.~Easson,
JCAP {\bf 0312}, 004 (2003)
[arXiv:hep-th/0301138]. 
  
\bibitem{reheat1}
N.~Barnaby, C.~P.~Burgess and J.~M.~Cline,
JCAP {\bf 0504}, 007 (2005)
[arXiv:hep-th/0412040];\\
L.~Kofman and P.~Yi,
Phys.\ Rev.\  D {\bf 72}, 106001 (2005)
[arXiv:hep-th/0507257];\\
A.~R.~Frey, A.~Mazumdar and R.~Myers,
Phys.\ Rev.\  D {\bf 73}, 026003 (2006)
[arXiv:hep-th/0508139];\\
D.~Chialva, G.~Shiu and B.~Underwood,
JHEP {\bf 0601}, 014 (2006)
[arXiv:hep-th/0508229].

\bibitem{reheat2}
S.~Mukohyama,
``Reheating a multi-throat universe by brane motion,''
arXiv:0706.3214 [hep-th].
  
\bibitem{gkp}
S.~B.~Giddings, S.~Kachru and J.~Polchinski,
Phys.\ Rev.\  D {\bf 66}, 106006 (2002)
[arXiv:hep-th/0105097].

\bibitem{ksgeom}
I.~R.~Klebanov and M.~J.~Strassler,
JHEP {\bf 0008}, 052 (2000)
[arXiv:hep-th/0007191].
  
\bibitem{ktgeom}
I.~R.~Klebanov and A.~A.~Tseytlin,
Nucl.\ Phys.\  B {\bf 578}, 123 (2000)
[arXiv:hep-th/0002159].
  

\bibitem{warpedtip}
S.~Kecskemeti, J.~Maiden, G.~Shiu and B.~Underwood,
JHEP {\bf 0609} (2006) 076
[arXiv:hep-th/0605189].\\
G.~Shiu and B.~Underwood,
Phys.\ Rev.\ Lett.\  {\bf 98} (2007) 051301
[arXiv:hep-th/0610151].

\bibitem{wrapped}
T.~Kobayashi, S.~Mukohyama and S.~Kinoshita,
``Constraints on Wrapped DBI Inflation in a Warped Throat,''
arXiv:0708.4285 [hep-th].\\
M.~Becker, L.~Leblond and S.~E.~Shandera,
``Inflation from Wrapped Branes,''
arXiv:0709.1170 [hep-th].
  
\bibitem{Thomas:2007sj}
S.~Thomas and J.~Ward,
Phys.\ Rev.\  D {\bf 76} (2007) 023509
[arXiv:hep-th/0702229].

\bibitem{HJ}
D.~S.~Salopek and J.~R.~Bond,
Phys.\ Rev.\  D {\bf 42} (1990) 3936.\\
W.~H.~Kinney,
Phys.\ Rev.\  D {\bf 56} (1997) 2002
[arXiv:hep-ph/9702427].

\bibitem{st}
E.~Silverstein and D.~Tong,
Phys.\ Rev.\ D {\bf 70}, 103505 (2004)
[arXiv:hep-th/0310221].

\bibitem{multiinfl}
C.~P.~Burgess, R.~Easther, A.~Mazumdar, D.~F.~Mota and T.~Multamaki,
JHEP {\bf 0505}, 067 (2005)
[arXiv:hep-th/0501125].
 
\bibitem{hko}
C.~P.~Herzog, I.~R.~Klebanov and P.~Ouyang,
``Remarks on the warped deformed conifold,''
arXiv:hep-th/0108101.
  
\bibitem{garmuk}
J.~Garriga and V.~F.~Mukhanov,
Phys.\ Lett.\ B {\bf 458} (1999) 219
[arXiv:hep-th/9904176].

\bibitem{AST}
M.~Alishahiha, E.~Silverstein and D.~Tong,
Phys.\ Rev.\  D {\bf 70}, 123505 (2004)
[arXiv:hep-th/0404084].

\bibitem{nongauss}
D.~Seery and J.~E.~Lidsey,
JCAP {\bf 0509} (2005) 011
[arXiv:astro-ph/0506056].\\
X.~Chen, M.~x.~Huang, S.~Kachru and G.~Shiu,
JCAP {\bf 0701} (2007) 002
[arXiv:hep-th/0605045].\\
T.~Battefeld and R.~Easther,
JCAP {\bf 0703}, 020 (2007)
[arXiv:astro-ph/0610296].

\bibitem{wands}
C.~Gordon, D.~Wands, B.~A.~Bassett and R.~Maartens,
Phys.\ Rev.\  D {\bf 63} (2001) 023506
[arXiv:astro-ph/0009131].


\bibitem{wandsrev}
D.~Wands,
``Multiple field inflation,''
arXiv:astro-ph/0702187; \\
  B.~A.~Bassett, S.~Tsujikawa and D.~Wands,
  Rev.\ Mod.\ Phys.\  {\bf 78} (2006) 537
  [arXiv:astro-ph/0507632].


\bibitem{2fieldpert}
Z.~Lalak, D.~Langlois, S.~Pokorski and K.~Turzynski,
``Curvature and isocurvature perturbations in two-field inflation,''
arXiv:0704.0212 [hep-th].
\bibitem{Panda:2007ie}
S.~Panda, M.~Sami and S.~Tsujikawa,
``Prospects of inflation in delicate D-brane cosmology,''
arXiv:0707.2848 [hep-th].

\bibitem{finelli}
F.~Di Marco, F.~Finelli and R.~Brandenberger,
Phys.\ Rev.\  D {\bf 67} (2003) 063512
[arXiv:astro-ph/0211276].

\bibitem{MFB}
V.~F.~Mukhanov, H.~A.~Feldman and R.~H.~Brandenberger,
Phys.\ Rept.\  {\bf 215}, 203 (1992).

\bibitem{constraints}
D.~Baumann and L.~McAllister,
Phys.\ Rev.\  D {\bf 75} (2007) 123508
[arXiv:hep-th/0610285];\\
R.~Bean, S.~E.~Shandera, S.~H.~Tye and J.~Xu,
JCAP {\bf 0705} (2007) 004
[arXiv:hep-th/0702107];\\
J.~E.~Lidsey and I.~Huston,
JCAP {\bf 0707} (2007) 002
[arXiv:0705.0240 [hep-th]];\\
H.~V.~Peiris, D.~Baumann, B.~Friedman and A.~Cooray,
``Phenomenology of D-Brane Inflation with General Speed of Sound,''
arXiv:0706.1240 [astro-ph].

\bibitem{Lev}
L.~Kofman and S.~Mukohyama,
``Rapid roll Inflation with Conformal Coupling,''
arXiv:0709.1952 [hep-th].

\end{thebibliography}
\end{document}